\documentclass[Physsubmission, Phys]{SciPost}
\hypersetup{
    colorlinks,
    linkcolor={red!50!black},
    citecolor={blue!50!black},
    urlcolor={blue!80!black}
}

\usepackage[bitstream-charter]{mathdesign}
\usepackage{cleveref}
\DeclareSymbolFont{usualmathcal}{OMS}{cmsy}{m}{n}
\DeclareSymbolFontAlphabet{\mathcal}{usualmathcal}

\begin{document}

% TODO: write your article's title here.
% The article title is centered, Large boldface, and should fit in two lines
\begin{center}{\Large \textbf{
      Monte Carlo Event Generator updates, \\ for $\tau$ pair events at Belle II energies
}}\end{center}

% TODO: write the author list here. Use initials + surname format.
% Separate subsequent authors by a comma, omit comma at the end of the list.
% Mark the corresponding author with a superscript *.
\begin{center}
Sw.~Banerjee\textsuperscript{1},
D.~Biswas\textsuperscript{1},
T.~Przedzinski\textsuperscript{2} and
Z.~Was\textsuperscript{3$\star$}
\end{center}

% TODO: write all affiliations here.
% Format: institute, city, country
\begin{center}
{\bf 1} University of Louisville, Louisville, Kentucky, 40292, USA
\\
{\bf 2} Institute of Physics, Jagellonian University, 30-348 Krakow, Lojasiewicza 11, Poland
\\
{\bf 3} Institute of Nuclear Physics, Polish Academy of Sciences, PL-31342 Krakow
\\
% TODO: provide email address of corresponding author
* Speaker
\end{center}

% For convenience during refereeing (optional),
% you can turn on line numbers by uncommenting the next line:
%\linenumbers
% You should run LaTeX twice in order for the line numbers to appear.

\definecolor{palegray}{gray}{0.95}
\begin{center}
\colorbox{palegray}{
%  \begin{tabular}{rr}
 % \begin{minipage}{0.1\textwidth}
%    \includegraphics[width=22mm]{Logo-DIS2021.png}
%  \end{minipage}
%  &
  \begin{minipage}{0.95\textwidth}
    \begin{center}
    {\it  16th International Workshop on Tau Lepton Physics (TAU2021),}\\
    {\it September 27 – October 1, 2021} \\
    \doi{10.21468/SciPostPhysProc.?}\\
    \end{center}
  \end{minipage}
%\end{tabular}
}
\end{center}

\begin{center}
Preprint IFJPAN-IV-2021-18,  November 2021
%\today
\end{center}

\vspace*{-1cm}

\section*{Abstract}
{\bf
% TODO: write your abstract here.
The Monte Carlo for lepton pair production and $\tau$ decays consist
of {\tt KKMC} for lepton pair production, {\tt tauola} for $\tau$ lepton decays and
{\tt photos} for radiative corrections in decays.

An effort for adaptation of the system for precision data to be collected
at {\tt Belle II} experiment lead to extension of phase space generation modules
both in {\tt photos} and {\tt tauola} to enable decays and/or radiative corrections
 of additional light lepton pairs. The phase-space and matrix
element parts of the programs are separated, that is why extension to
processes where lepton pair is produced through narrow resonances,
like dark photon, was straight forward.

In the present version of {\tt tauola},
the list of $\tau$ decay channels is enriched with multitude of exotic ones, useful for
searches of new physics. The hadronic currents parametrization of main decay channels
is prepared for basic simulation in the experiment. The basis for future
work on precise fits of hadronic currents including Machine Learning
is retained, but development of necessary software solutions is left for
the forthcoming years.

Presented programs versions are available in stand-alone format from GitLab or through the {\tt basf2} system of
Belle II software. Official distribution web pages, documenting programs tests, are retained but not
necessarily up to date.}

% TODO: include a table of contents (optional)
% Guideline: if your paper is longer that 6 pages, include a TOC
% To remove the TOC, simply cut the following block
\vspace{10pt}
\noindent\rule{\textwidth}{1pt}
\tableofcontents\thispagestyle{fancy}
\noindent\rule{\textwidth}{1pt}
\vspace{10pt}

\section{Introduction}
\label{intro}

The {\tt tauola} package
\cite{Jadach:1990mz,Jezabek:1991qp,Jadach:1993hs,Golonka:2003xt} for simulation
of $\tau$-lepton decays and
{\tt photos} \cite{Barberio:1990ms,Barberio:1994qi,Golonka:2005pn} for simulation of QED radiative corrections
in decays, are computing
projects with a rather long history. Written and maintained by
well-defined principal authors, they nonetheless migrated into a wide range
of applications where they became essential ingredients of
complicated simulation chains. In the following, we shall concentrate
on the version of the programs which are prepared for installation
in {\tt basf2} software of the Belle II experiment. The following programs are installed in the system:
  (i) {\tt KKMC} for the $\tau$ lepton production process
  $e^-e^+ \to \tau^-\tau^+ n\gamma$
  (ii)  {\tt tauola} for $\tau$ lepton decays,
  (iii) {\tt photos} for bremsstrahlung in decays of particles and resonances,
  (iv)  {\tt photospp}, the C++ version of {\tt photos}, which at present
  is used only for supplementing events with lepton pairs, produced through
  virtual careers of the electroweak interaction in the Standard Model or through New Physics processes. 

Our presentation is organized as follows:
Section 2 is devoted to technical changes prepared in Ref.~\cite{Antropov:2019ald} for {\tt tauola} and earlier in Ref.~\cite{Davidson:2010ew}
for {\tt photos}.  
These  extensions  of phase space generators, enable not only
full implementation of bremsstrahlung-like processes where virtual photon decay
into a pair of leptons, but also the possibilities of emitting dark photon or dark scalars,
now introduced into {\tt tauola} and {\tt photospp}.
In Section 3 we concentrate on   {\tt photos} Monte Carlo for
radiative corrections in decays. The new version of the program 
features emissions of light lepton pairs, which may originate from virtual
photon or from dark photons or scalars. Numerical tests of dark scalar implementation
into {\tt KKMC} \cite{Jadach:1999vf} $e^-e^+ \to \tau^-\tau^+ n\gamma$ event samples are presented.
Section 4  is devoted to the discussion of initialization for 
 {\tt tauola} hadronic currents and new decay channels as prepared for {\tt basf2} of Belle II.
This point was already announced in  \cite{Was:2014zma} and presented in \cite{Chrzaszcz:2016fte}, that is why, we will address only those points which may be important for the future users.
Section 5 summarizes this proceeding.
%is devoted to  applications of {\tt TAUOLA}
%  for hard processes with final state $\tau$ leptons. In particular for
%observable construction and evaluation of its sensitivity.
% The techniques,
%Machine Learning (ML),   were found to be useful in analyses of HEP data.
% In this context we discuss  {\tt TauSpinner} algorithm, which was found useful for evaluation 
%of observable for Higgs boson parity measurement. We mention 
% other applications or tests; in particular in the domain of 
%algorithm of calculating spin states of $\tau$ pairs in events where
%high $p_T$ jets are present in $pp$ collisions. The following two sections provide
%examples for applications in the domain of precision measurements and Higgs CP parity
%evaluation at LHC.
% Summary Section 7, closes the presentation.

%%%%%%%%%%%%%%%%%%%%%%%%%%%%%%%%%%%%%%%%%%%%%%%%%%%%%%%%%%%%%%%%%%%%%%%%%%%%%%%%%%%%%%%%%%%%%
\section{ Technical aspects: phase space presamplers}
%%%%%%%%%%%%%%%%%%%%%%%%%%%%%%%%%%%%%%%%%%%%%%%%%%%%%%%%%%%%%%%%%%%%%%%%%%%%%%%%%%%%%%%%%%%%%
From a technical point of view, the most important update was the introduction of
phase space presamplers for lepton pairs originating from virtual photon
or narrow exotic resonances, either vector or scalar in nature. Both {\tt tauola} and {\tt photos} rely
on exact phase space parametrization. In principle, results of simulations
also depend on parameterization of matrix elements and/or form factors used, but if presamplers of phase-space are not appropriate,
efficiency of generation is poor and in extreme cases the distributions may be unreliable.
Usually, the weight monitoring functionalities point to the difficulties.
In fact, the necessity of changes can be identified from inspection of matrix
elements to be installed. Details for the case of lepton pairs can be found in
Ref.~\cite{Antropov:2019ald} and are not repeated here. What is
important, is that phase space parametrization in both {\tt tauola} and {\tt photospp} now have presamplers for
virtual photon or narrow resonances decaying to lepton pairs.

Once matrix element installed and distributions generated, it is necessary to
perform tests. Semi-analytical, or analytical results for pair emissions are rare,
but nonetheless were reported in Ref.~\cite{Antropov:2019ald} and 
\cite{Was:2014zma}. These numerical results are mostly about final state rates
and rely on eikonal approximations for matrix elements.
%[{\it Some more references here, who volunteer to read and add?}]
Pair emission can also be generated with other generators. That was used
for test already long time ago. Automated comparison package {\tt MC-TESTER}
\cite{Davidson:2008ma} was used to construct histograms and compare results from {\tt KORALW}
\cite{koralw:1998} with those from {\tt photos}  in tandem with {\tt KKMC}.
Such  tests are very helpful. We have used them in study of CP sensitive observables
in $H \to \tau \tau $ channel~\cite{Lasocha:2020ctd},
where events simulated with {\tt tauola universal interface} \cite{Davidson:2010rw}
are compared with those including  parity effects introduced
with the {\tt tauspinner} algorithm \cite{Przedzinski:2018ett}.
For $\tau \tau $ {\it jet jet} such test and comparisons  with {\tt MadGraph}~\cite{Alwall:2014hca}
are reported in \cite{Bahmani:2017wbm}. Now events simulated with  {\tt photos} 
were compared with events simulated with {\tt MadGraph}  for the $e^-e^+ \to \tau^- \tau^+ X\; (X \to  l \bar{l})$ process,
where X is an exotic particle motivated by new physics models,
and can be vector $Z^\prime$ \cite{Shuve:2014doa, Altmannshofer:2016jzy} or dark scalar \cite{Batell:2016ove}.
The spin state of $\tau$ flips when X is scalar,
which needs to be taken into account in proper simulation of $\tau-\tau$ spin correlations.
For these developments,  distributions prepared with {\tt MC-TESTER}
were obtained, and used in the development of {\tt tauola} and {\tt photospp}.
%as addressed in the following section.

%%%%%%%%%%%%%%%%%%%%%%%%%%%%%%%%%%%%%%%%%%%%%%%%%%%%%%%%%%%%%%%%%%%%%%%%%%%%%%%%%%%%%%%%%%%%%
\section{{\tt PHOTOS} Monte Carlo for bremsstrahlung and lepton pair emission}
%\newline its systematic uncertainties}
%\def\CCol{{\tt SANC}}
%%%%%%%%%%%%%%%%%%%%%%%%%%%%%%%%%%%%%%%%%%%%%%%%%%%%%%%%%%%%%%%%%%%%%%%%%%%%%%%%%%%%%%%%%%%%%

\begin{figure}[!h]
  \begin{center}
    \includegraphics[width=0.49\textwidth]{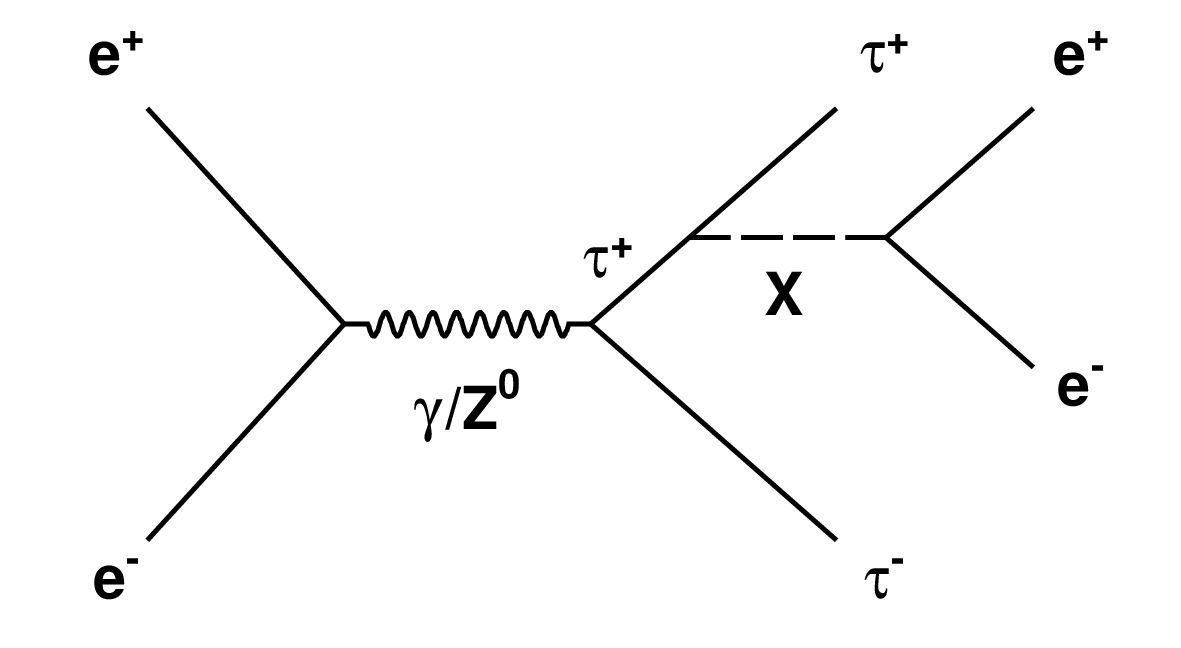}  
    \includegraphics[width=0.49\textwidth]{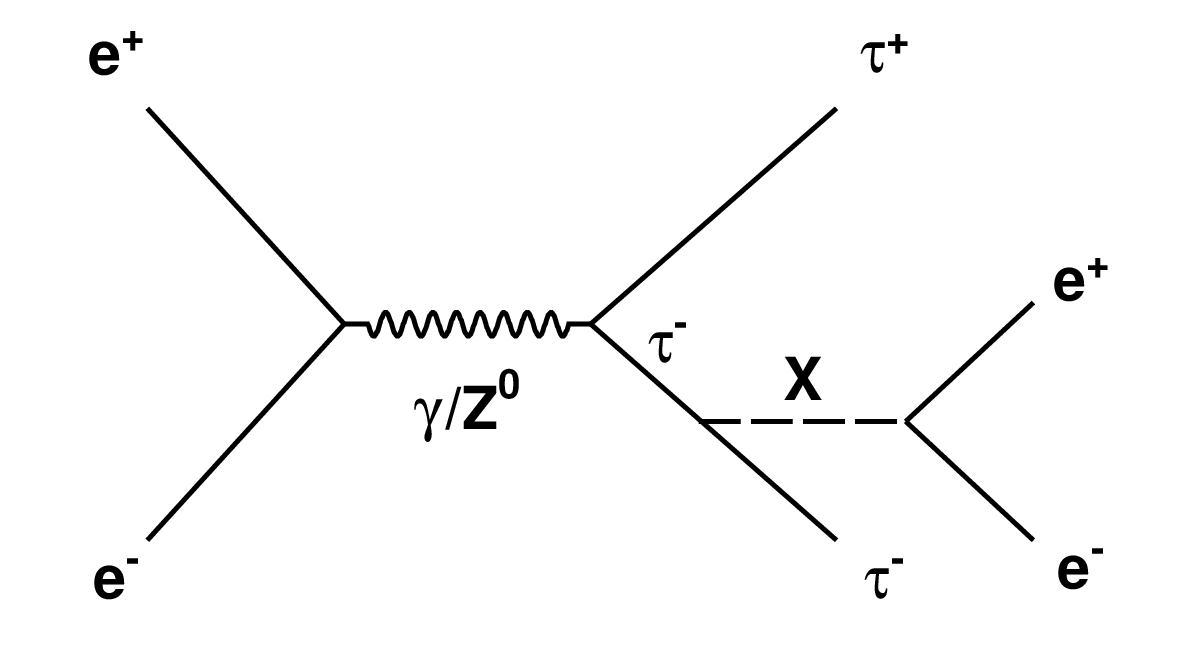}  
    \caption{Feynman diagrams for $e^-e^+ \to \tau^- \tau^+ X (\to e^- e^+)$,
      used for preparation of {\tt photospp} emission kernel.}
%  as installed in {\tt MadGraph}. For our applications dominant contributions which can
%  be used together with bremsstrahlung,  is used. This approximation
%  is validated in Figs.~\ref{fig:ee} and~\ref{fig:mumu}. It also opens up the gate for simultaneous inclusion of large QED effects.
%        {\it .}}
\label{fig:Feyn}
\end{center}
\end{figure}

Numerical tests for pair emission algorithm are  published \cite{Antropov:2017bed}, 
and following updates to the program presented in \cite{Davidson:2010ew}.
This update opened up new possibilities, in particular,
generation of lepton pairs in the process
$e^-e^+ \to \tau^-\tau^+ n\gamma$ with $\tau$ decays and implementation into final state of
an exotic particle X motivated by new physics.
There are two reasons, why matrix elements of refs  \cite{Shuve:2014doa,Altmannshofer:2016jzy,Batell:2016ove} could not be
used directly. First is to preserve modularity of {\tt photos} Monte Carlo
design. The second is because the matrix element form must enable its interpolation for use when additional bremsstrahlung photons are present. That is why
a factorized form, with the emission factor similar to the eikonal one, was necessary to be devised and checked for the process shown in Fig.~\ref{fig:Feyn}.

%\footnote{
%Let us provide some details.
In QED, the eikonal factor is the difference between
  the amplitude of some scattering process and the one when additional soft photon
  is present. If there are many photons in final states and all are at
  small energies, products of such factors, are added.
  In the soft photon limit, each such factor depends only on momenta of outgoing and incoming
  charged particles and on the momentum of the photon. The resulting
  formula is quite simple $\sum_{\rm{charged~particles}} Q_i \frac{2p^\mu_i}{2p_i\cdot k}$
  and well known. What is important is that it can be expressed with four
  momenta and can be used all over the phase space.
  One should not forget  it is the zeroth level 
  approximation,  and thus, corrections obtained from matrix element calculations
  are necessary. 
  This is an essential element of exclusive exponentiation for photon
  generation in  {\tt KKMC} \cite{Jadach:1999vf} or {\tt photos}.
  Hard
  photon configuration corrections in general depend
  on momenta of all photons in a particular event.

  The QED matrix elements for configurations with additional lepton pairs has
  a similar form if lepton pair virtuality and energies are small. In this
  case, the emission factor is process independent, and has a form which
  can be used all over the phase space. In {\tt photospp} variant of \cref{eq1}
  from \cite{Jadach:1993wk} is used.
  
  For configurations where lepton pairs from decay of exotic scalars
  or vector particles are present, approximate matrix elements were derived,
  following educated guesses.
  The approximations were then validated with {\tt MadGraph} simulation~\cite{Alwall:2014hca}
 samples.
  The best of  several variants was chosen.
  This opened up the gateway for  simultaneous inclusion of large QED effects, e.g. ISR as implemented in {\tt KKMC}. ISR effects 
  were incorporated in {\tt MadGraph} simulation using the recipe from \cite{Li:2018qnh}.
  
\begin{figure}[!h]
  \begin{center}
  \includegraphics[width=0.48\textwidth]{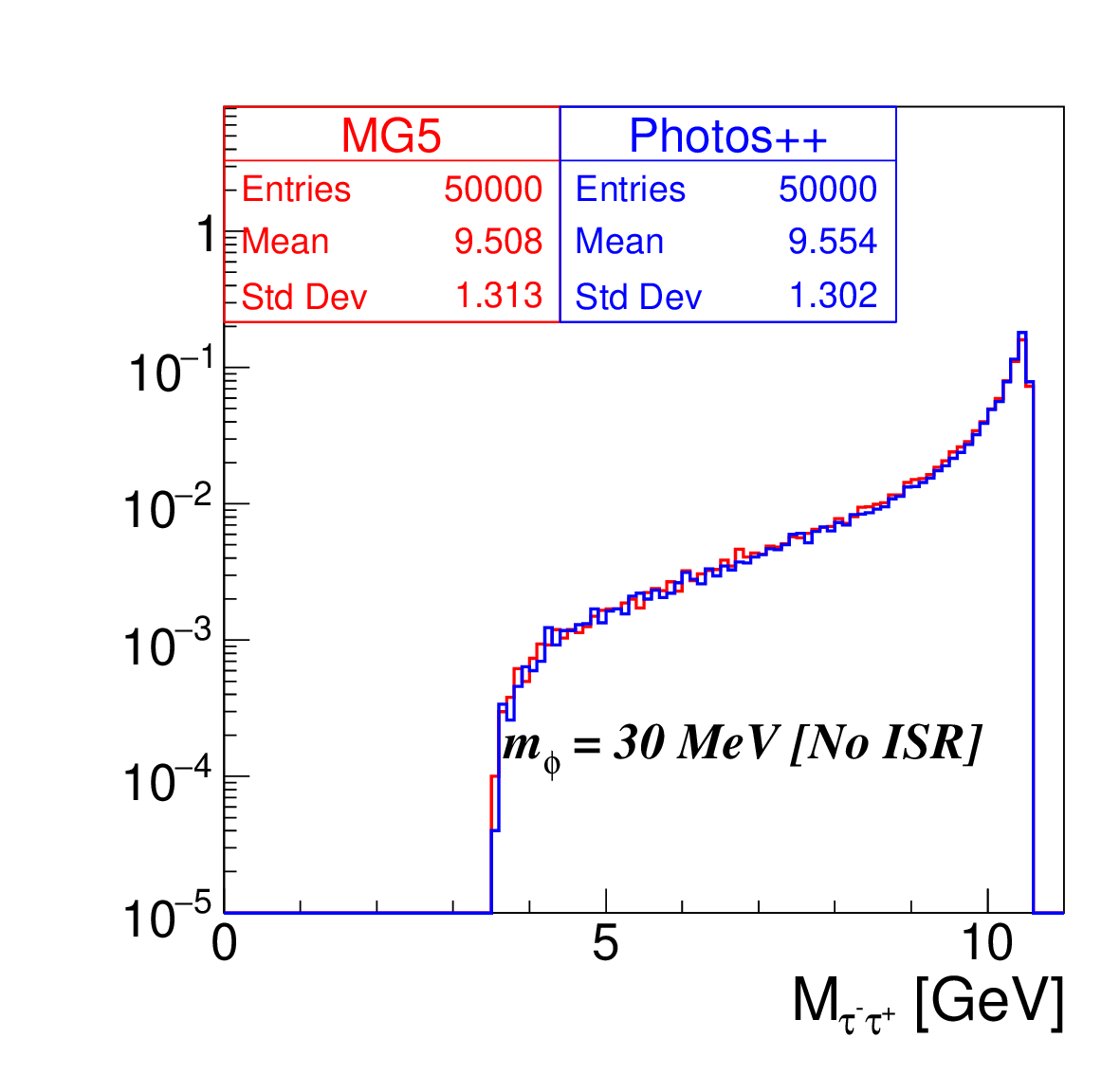}
  \includegraphics[width=0.48\textwidth]{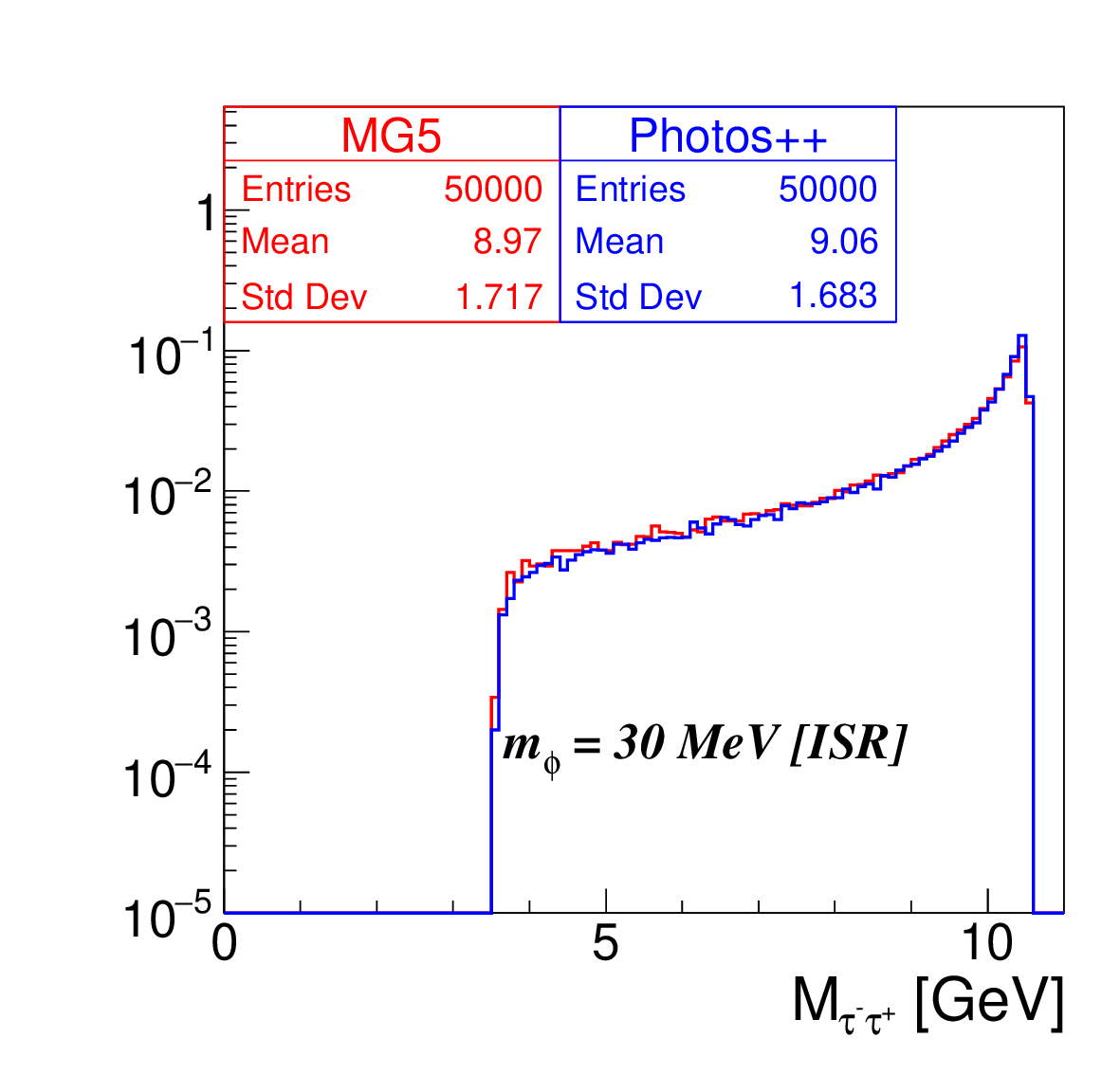}\\
  \includegraphics[width=0.48\textwidth]{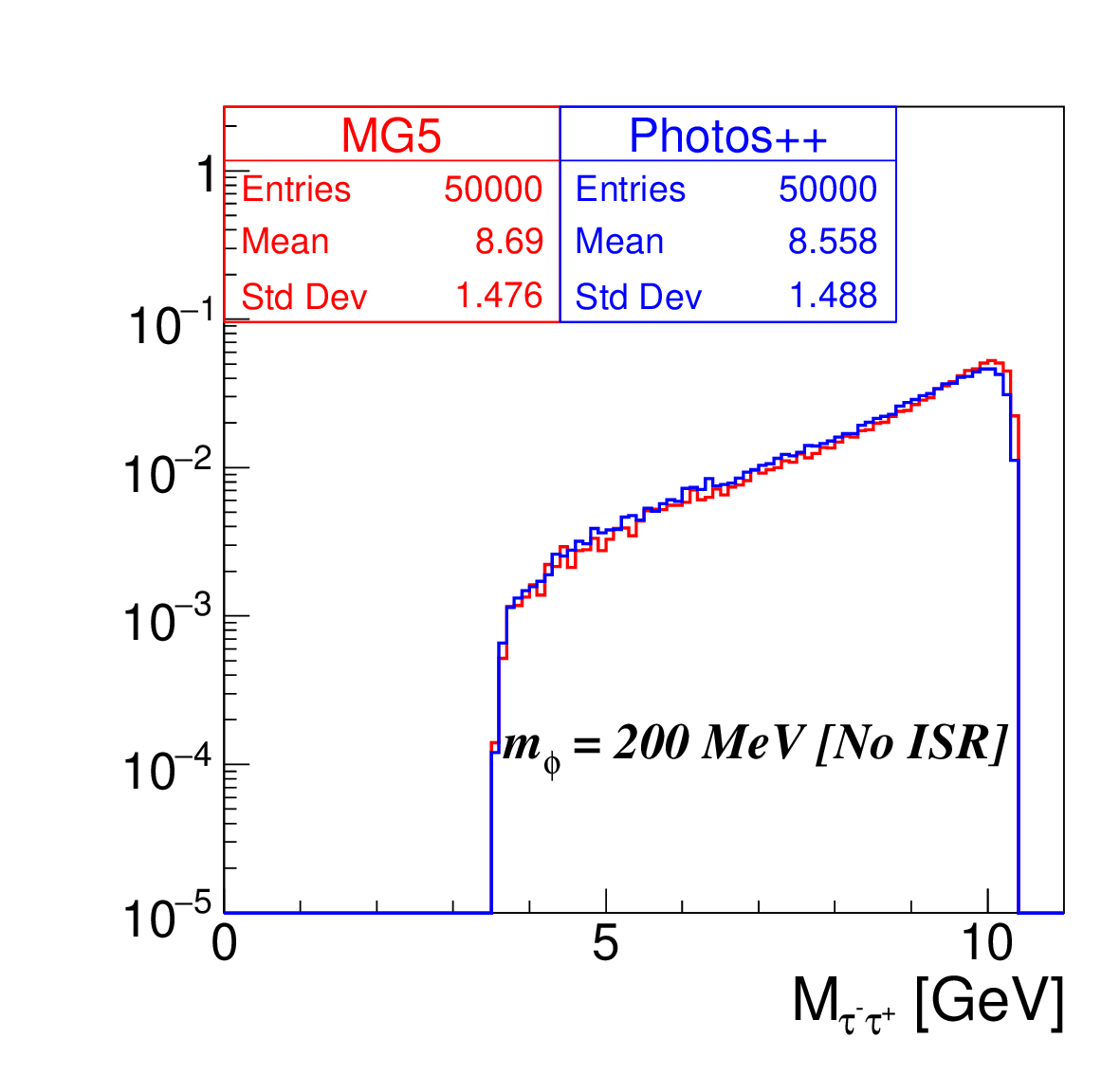}
  \includegraphics[width=0.48\textwidth]{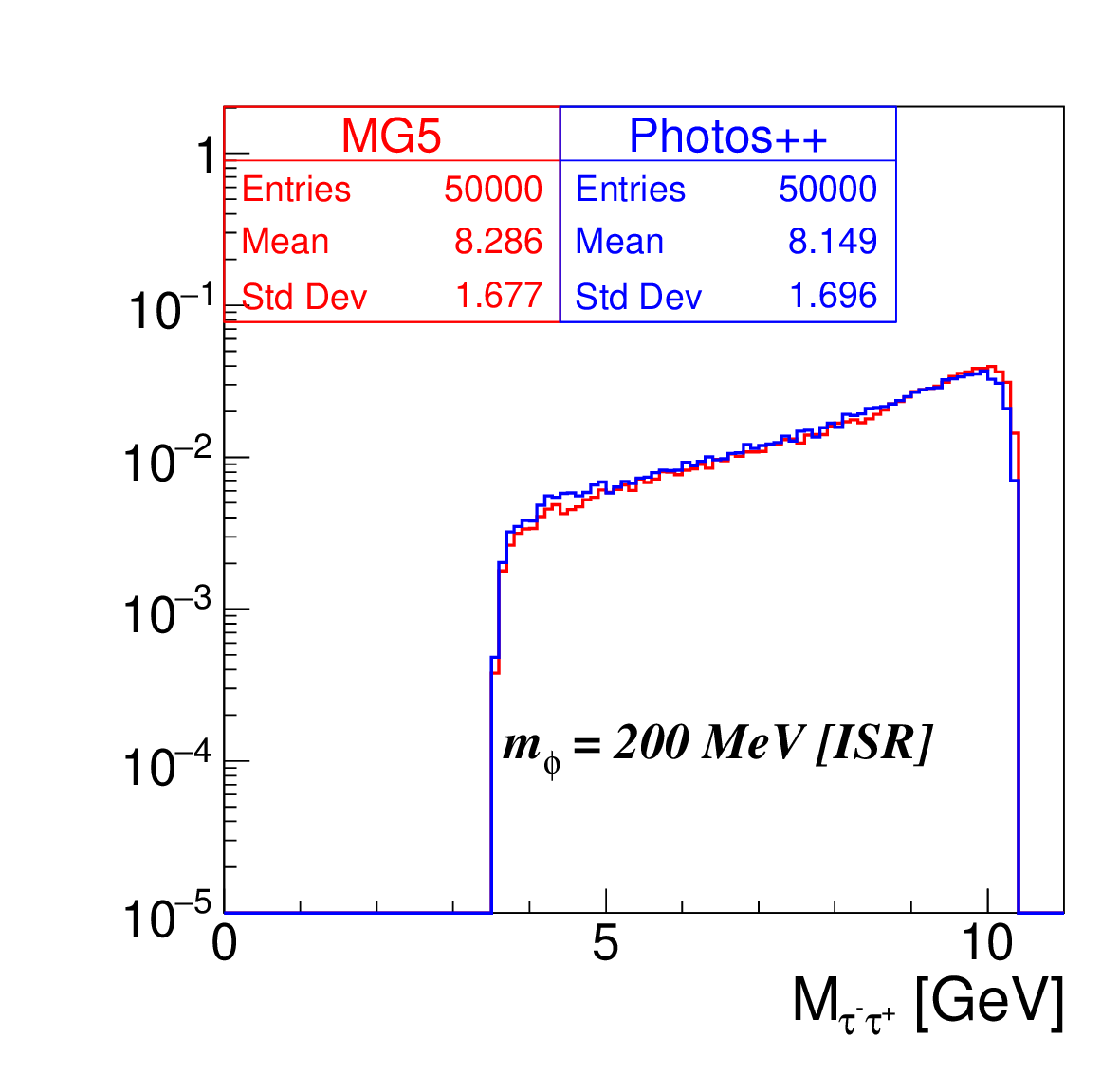}
  \caption{$e^-e^+ \to \tau^- \tau^+ \phi_{\rm{Dark~Scalar}} (\to e^-e^+)$}\label{fig:ee}
  \end{center}
 \end{figure}

\begin{figure}[!h]
  \begin{center}
\includegraphics[width=0.48\textwidth]{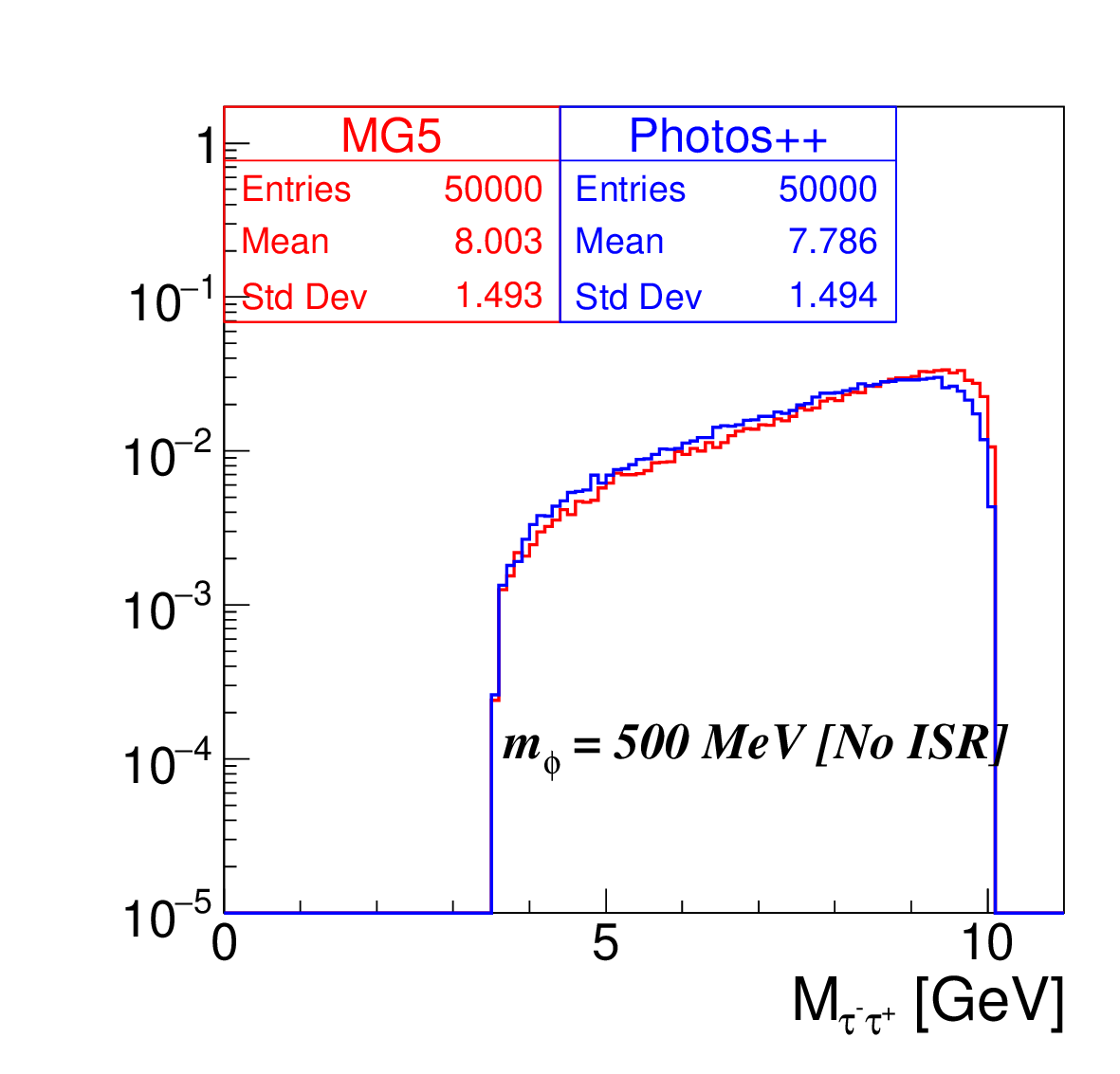}
\includegraphics[width=0.48\textwidth]{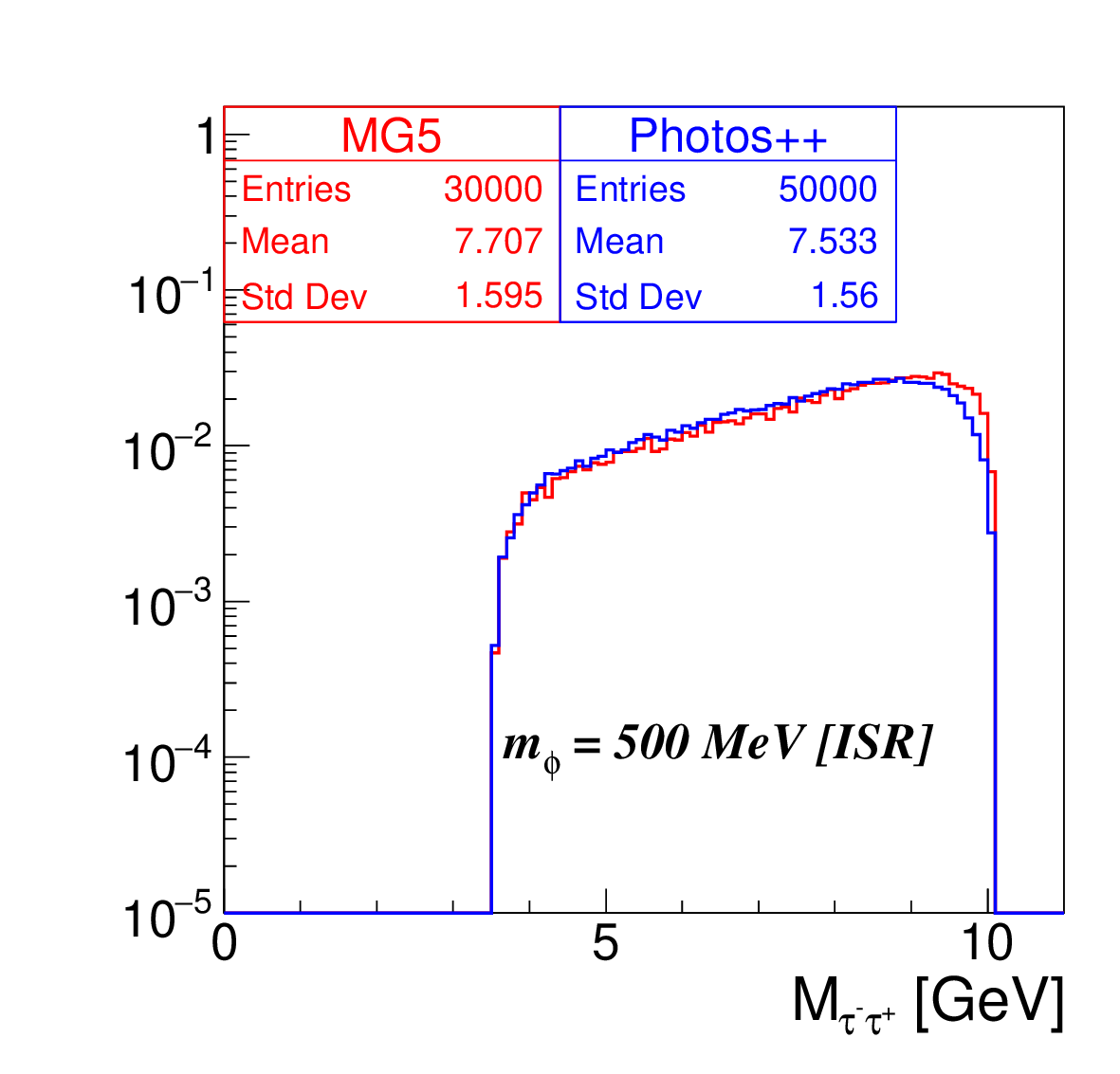}\\
\includegraphics[width=0.48\textwidth]{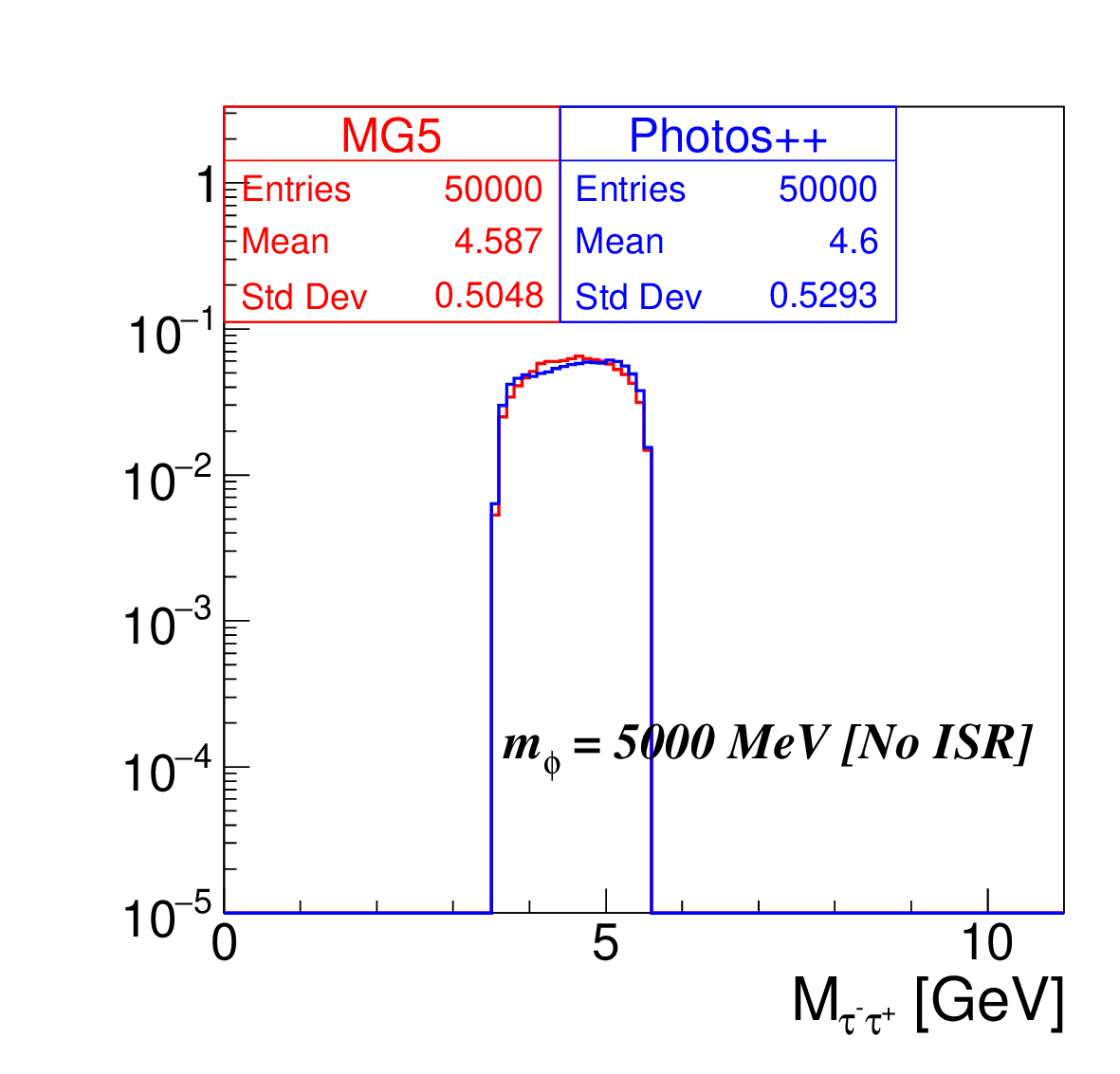}
\includegraphics[width=0.48\textwidth]{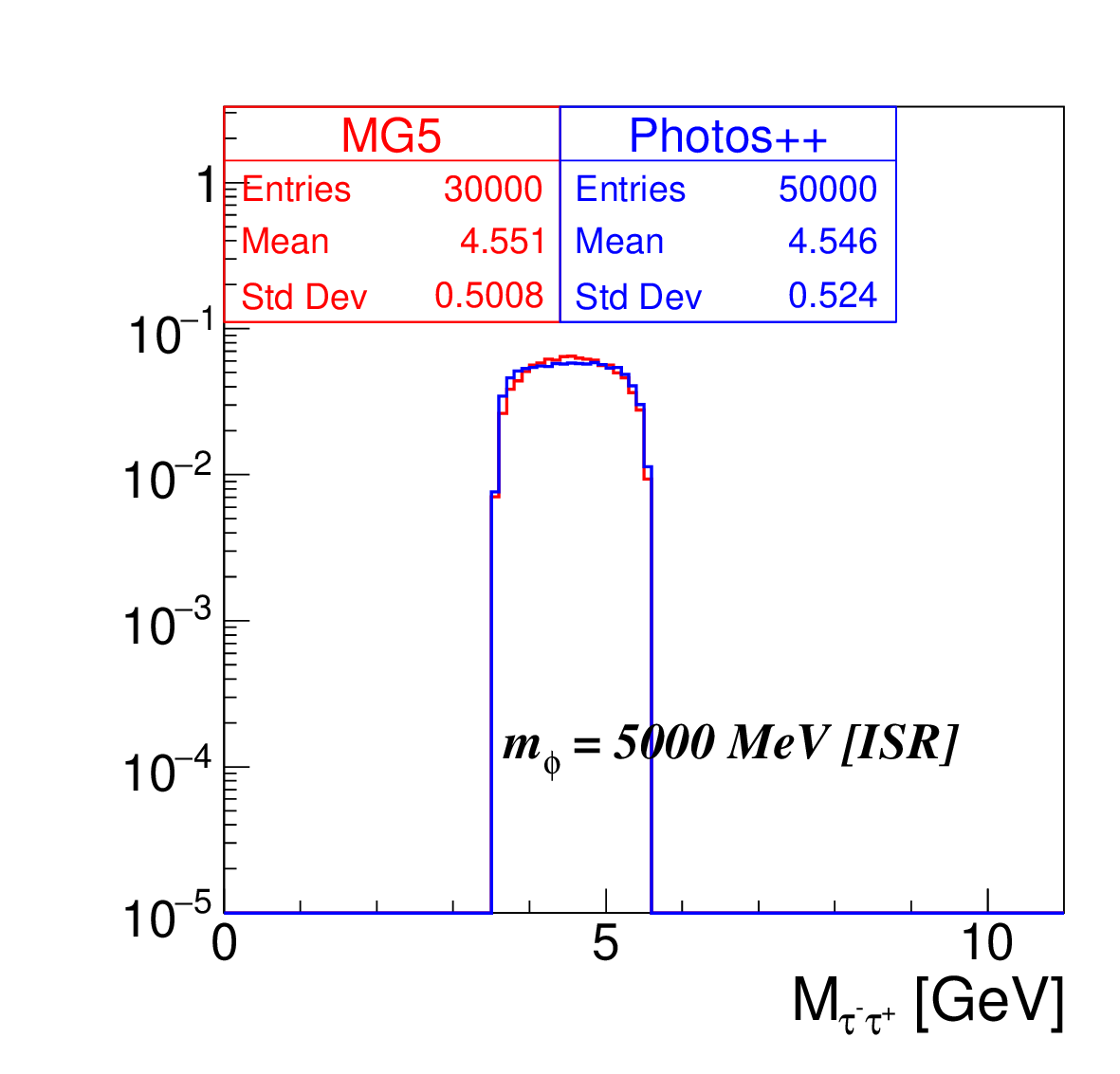}
\caption{ $e^-e^+ \to \tau^- \tau^+ \phi_{\rm{Dark~Scalar}} (\to \mu^-\mu^+)$ } \label{fig:mumu}
  \end{center}
\end{figure}

In fact, not only test with {\tt MadGraph} simulation~\cite{Alwall:2014hca} samples were necessary. Several iterations of {\tt photos}
matrix elements were performed  to achieve better a simulation  tool.
%}
Validations and choices were performed  with the help of {\tt MC-TESTER}
Shape Difference Parameter.  Final validation of  {\tt photos} was the check
on the distributions
of the recoil mass of the $\tau$-pair system
for the process $e^-e^+ \to \tau^- \tau^+ \phi_{\rm{Dark~Scalar}}$,
where the $\phi_{\rm{Dark~Scalar}}$ decays into a pair of oppositely charged electrons or muons,
as shown in Figs.~\ref{fig:ee} and~\ref{fig:mumu}.

The  {\tt photospp} version of {\tt photos} is implemented in C, its external parts, the interfaces are implemented
in C++.
In general,  it features both photon and lepton pair emissions in decays of resonances or particles.
    In particular the emitted lepton pairs  can be due to  QED interactions or due to dark photon or dark scalars.

    The {\tt photospp} generation can be combined with {\tt KKMC+tauola+photos.f} one. For the sake of rapid developments,
    {\tt photospp} was not used for photon emissions. This enables more flexibility and some testing could be spared. 
  The {\tt photospp} is used for construction of  final states of the type
  $Z/\gamma^* \to  \tau \tau X(scalar/pseudoscalar/vector/\gamma^*)$ with usual $\tau$ decays and
  with full $\tau \tau$ spin correlations and with  $X \to ee (\mu\mu)$ decays too.

  Let us indicate some technical foundations in the following.
  The partial width of the decay of a resonance or particle to an intermediate
  state of n final state particles, is given by the following formula:
\begin{equation}
d\Gamma = \frac{1}{2M}|\bar{M}_n|^{2} dLips_n = \frac{1}{2M}|\bar{M}_n|^{2} {J_n} \prod_{j=1}^{k_n} dx_j,
\label{eq1}
\end{equation}
where $M$ denotes the mass of the decaying object, $\bar{M}_n$ denotes the decay matrix element and
$dLips_n$ denotes the Lorentz invariant phase space integration element. The $dLips_n$
can be expressed, after change of variables, as a product of integration elements
$dx_j$, of the $x_j$ real integration variables in the [0,1] range. Finally
$J_n$ denotes Jacobian of the variable change.
This formula is used in {\tt photos}. Formally speaking, it is used twice, first for
the n-body decay of the input particle, where the
$x_j$ vector is recovered from the event to be modified by {\tt photos}.
Then the set of $\{x_j\}$  is supplemented with additional $x_j'$ to complement the set of
coordinates necessary to parametrize n+1 or n+2 body phase space. These are
finally used for construction of event where particles are added.
That is the general idea, which explains why the algorithm can perform without
approximations for phase space, and why the full phase space coverage can be
guaranteed. Explanations of complications due to matrix element with
collinear/soft or narrow resonance approximations are now omitted.
We address the reader to
Refs.~\cite{Davidson:2010ew,Nanava:2006vv} for further details.

%%%%%%%%%%%%%%%%%%%%%%%%%%%%%%%%%%%%%%%%%%%%%%%%%%%%%%%%%%%%%%%%%%%%%%%%%%%%%%%%%%%%%%%%%%%%%
\section{ Status of  {\tt tauola} Monte Carlo initializations}
%%%%%%%%%%%%%%%%%%%%%%%%%%%%%%%%%%%%%%%%%%%%%%%%%%%%%%%%%%%%%%%%%%%%%%%%%%%%%%%%%%%%%%%%%%%%%
We do not aim at repeating, this what was already documented
in Ref.~\cite{Antropov:2017bed} and in references therein,  but we
concentrate on recent changes, introduced for Belle II applications.

\begin{figure}[!h]
  \begin{center}
    \includegraphics[width=0.49\textwidth]{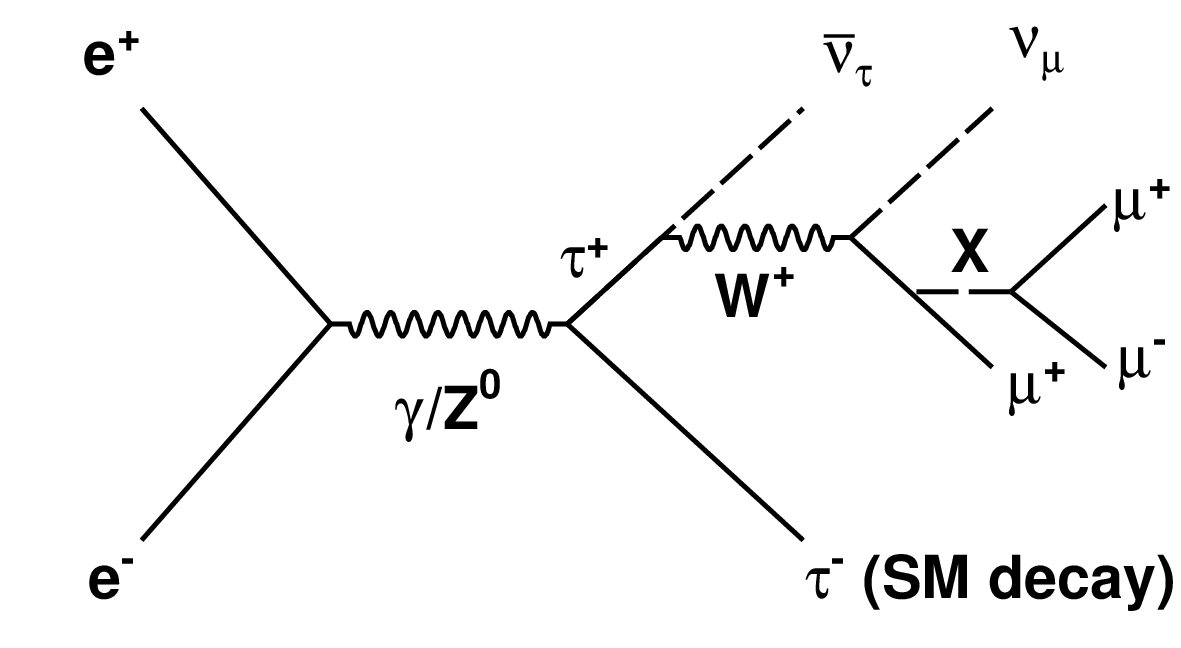}
   \caption{ Feynman diagram for $e^-e^+ \to \tau^-~({SM~decay})~ \tau^+ (\to \bar{\nu}_\tau \nu_\mu \mu^+ X (\to \mu^- \mu^+))$ process with  X emitted in $\tau^+$ decay, used in {\tt tauola}. }
   \label{fig:decay}
\end{center}
\end{figure}

For simulations in the Belle II software, many channels were prepared to establish
defaults:
 \begin{enumerate} 
\item
  Total number of decay channels:  278
\item
  2 body neutrinoless non SM decays: 58
\item
  3 body neutrinoless non SM decays: 46
\item
  Number of generic SM decay channels : 92, initialized with PDG Branching Fractions~\cite{PhysRevD.98.030001}.
  For matrix elements, choices from  older versions of parametrization
  were taken, except:
  \begin{enumerate}
  \item for high precision data obtained for $\tau^- \to \pi^- \pi^0 \nu_\tau$ decays~\cite{Fujikawa:2008ma} by the Belle collaboration,
    and for $\tau^- \to \pi^- \pi^+ \pi^- \nu_\tau$ decays by the BaBar collaboration using the Resonance Chiral Lagrangian initialization~\cite{Nugent:2013hxa}.
\item Theoretical uncertainty of models is worse than  quality
  of data. That is why, new, or alternative decay modes can be installed by the
  user with the help of specially prepared for that routines but must be validated by experimental data. User can choose
  decay product flavors, their masses, as well as matrix element or hadronic current and tune it to data in the future.
 Note that temporarily  this option  inactive, pending  necessary arrangements for basf2 library.
\end{enumerate} 
\item
  New decay modes with SM photons or Dark photons decaying to lepton pair with mass $\in$ $[50, 1500]$ MeV
  with matrix elements cross-validated with {\tt MadGraph}~\cite{Alwall:2014hca} for
  \begin{itemize}
  \item $\tau^- \to \nu_\tau \bar{\nu}_{\ell} \ell^- \ell^+ \ell^-$ decays.
    A typical Feynman diagram for such a process is shown
      in Fig.~\ref{fig:decay}.\\
    \vspace*{-.5cm}
    \item $\tau^- \to \nu_\tau \pi^- \ell^+ \ell^-$ decays.
   \end{itemize}
 \end{enumerate}

\section{  Summary }
%%%%%%%%%%%%%%%%%%%%%%%%%%%%%%%%%%%%%%%%%%%%%%%%%%%%%%%%%%%%%%%%%%%%%%%%%%%%%%%%%%%%%%%%%%%%%

%\begin{enumerate}
%\item
Final states of $\tau$ lepton pairs with bremmstrahlung photons
and dark scalar/photon (decaying to the lepton pair) were 
introduced for  study of $e^-e^+$ collisions. The $\tau$ pair production and
decay were introduced with {\tt KKMC} and  {\tt tauola}, dark scalar/photon
with {\tt photospp}.
% \item
Efforts on $\tau$ lepton decays have been focused on channels, their
matrix element and presamplers initializations.
An extended list of $\tau$ decay channels is now available in the {\tt basf2}
software for the Belle II collaboration, in particular New Physics
neutrinoless ones.
% \item

  Introductory steps for language change in {\tt tauola} are completed.
% \item
   Parts aimed for migration into C are localized. Parts more convenient
     for migration into C++ are 
     separated as well. Evolution is prepared to follow the same development path
     as was chosen for {\tt photos}; its C++ version is already  installed.
%   \item
     Also, C++ version, focused on FCC applications  of {\tt KKMC} exist.
     
    % Not published yet, but available through gitlab. 
%   \item
     Not much progress has been made for re-arrangements for fits of multidimensional signatures
     and ML solutions, but it is kept in mind, even if temporarily deactivated.
%   \item

     The speaker of this  talk (Z.W.) hopes to provide help with future developments of {\tt tauola},
     perhaps including reweighting techniques with  {\tt TauSpinner} weights,
     originally deployed for the study of spin-parity of the Higgs Boson,
     but this time applicable to the physics of  $\tau$ decays.
%\end{enumerate}

\section*{Acknowledgements}
{\small 
This project was supported in part from funds of Polish National
Science Centre under decisions DEC-2017/27/B/ST2/01391.
}

% TODO:
% Provide your bibliography here. You have two options:

% FIRST OPTION - write your entries here directly, following the example below, including Author(s), Title, Journal Ref. with year in parentheses at the end, followed by the DOI number.

% SECOND OPTION:
% Use your bibtex library
% \bibliographystyle{SciPost_bibstyle} % Include this style file here only if you are not using our template
%\bibliography{Tauola_interface_design.bib}

\nolinenumbers

\end{document}